\documentclass[fleqn,usenatbib]{mnras}

\usepackage{newtxtext,newtxmath}
\usepackage[T1]{fontenc}
\usepackage{ae,aecompl}

\usepackage{graphicx}	% Including figure files
\usepackage{amsmath}	% Advanced maths commands
\usepackage{color} 
\title[Fast-varying time lags in GRS 1915+105]{Fast-varying time lags in the Quasi-periodic Oscillation in GRS 1915+105}

\author[T.M. Belloni et al.]{
Tomaso M. Belloni$^{1,\dagger}$, %\thanks{E-mail: tomaso.belloni@inaf.it}
Mariano M\'endez$^{2}$\thanks{E-mail: mariano@astro.rug.nl},
Federico Garc\'{\i}a$^{3}$
and Dipankar Bhattacharya,$^{4,5}$
\\
$^{1}$INAF - Osservatorio Astronomico di Brera, via E. Bianchi 46, I-23807 Merate, Italy\\
$^{2}$Kapteyn Astronomical Institute, University of Groningen, PO BOX 800, NL-9700 AV Groningen, the Netherlands\\
$^{3}$Instituto Argentino de Radioastronom\'{\i}a (CCT La Plata, CONICET; CICPBA; UNLP), C.C.5, (1894) Villa Elisa, Buenos Aires, Argentina\\
$^{4}$Inter University Center for Astronomy and Astrophysics, Ganeshkhind, Post Bag 4, Pune 411007, India\\
$^{5}$Ashoka University, Dept. Of Physics, Sonipat, Haryana-131029, India\\
$^{\dagger}$Deceased
}

\date{Accepted XXX. Received YYY; in original form ZZZ}

\pubyear{2023}

\begin{document}
\label{firstpage}
\pagerange{\pageref{firstpage}--\pageref{lastpage}}
\maketitle

\begin{abstract}
The properties of sub-second time variability of the X-ray emission of the black-hole binary GRS 1915+105 are very complex and strictly connected to its patterns of variability observed on long time scales. A key aspect for determining the geometry of the accretion flow is the study of time lags between emission at different energies, as they are associated to key time scales of the system. In particular, it is important to examine the lags associated to the strong low-frequency Quasi-periodic Oscillations (QPOs), as the QPOs provide unambiguous special frequencies to sample the variability.
We have analyzed data from an observation with the AstroSat satellite, in which the frequency of the low-frequency QPO varies smoothly between 2.5 and 6.6 Hz on a time scale of $\sim$10 hours. The derived phase lags show the same properties and evolution of those observed on time scales of a few hundred days, indicating that changes in the system geometry can take place on times below one day. We fit selected energy spectra of the source and rms and phase-lag spectra of the QPO with a time-variable Comptonization model, as done previously to RossiXTE data of the same source, and find that indeed the derived parameters match those obtained for variations on much longer time scales.
\end{abstract}


%%%%%%%%%%%%%%%%%%%%%%%%%%%%%%%%%%%%%%%%%%%%%%%%%%

%%%%%%%%%%%%%%%%% BODY OF PAPER %%%%%%%%%%%%%%%%%%

\section{Introduction}
\label{sec:introduction}
The sub-second variability in the X-ray emission of black-hole binaries (BHBs) displays a large range of features, ranging from broad-band noise components to several types of Quasi-periodic Oscillations (QPOs) that can be observed on characteristic time scales from a few millihertz to hundreds of Hz (see Belloni \& Stella 2014, Belloni \&  Motta 2016, Ingram \& Motta 2020).
While high-frequency QPOs, with centroid frequencies above 40 Hz, are very rarely detected (Belloni, Sanna \& M\'endez 2012), their low-frequency counterparts, spanning frequencies typically in the $0.1-30$ Hz range, have been observed several times from almost all objects. These QPOs have been divided into three separate classes, called A, B and C (Wijnands \& van der Klis 1999, Remillard et al. 2002, Casella et al. 2004, Ingram \& Motta 2020), usually mutually exclusive although some cases of two different QPOs appearing at the same time have been observed (see e.g. Motta et al. 2012).
The most commonly observed are type-C QPOs, which in the Power Density Spectra (PDS) are associated to the presence of a strong band-limited noise that can be decomposed into a few components (Belloni \& Stella 2014). The centroid frequency of these QPOs has been interpreted as being associated to the Lense-Thirring frequency at a special radius in the accretion flow (Stella \& Vietri 1998). A more complex model based on the same physical frequency was later proposed (Ingram et al. 2009, see also Ingram \& Motta 2020). Additional models have been proposed (a review can be found in Belloni \& Stella 2014). Recently, a different model was introduced by Mastichiadis, Petropoulou \& Kylafis (2022), in which the type-C QPO is generated by the interaction between the corona and the accretion disk. In this model, the frequency of the oscillation is associated to accretion time scales in the interaction between the two physical components. 
All these models address the observed timing properties and are not directly linked to the energy spectra.

Frequency-dependent lags between different X-ray energy bands associated to aperiodic variability in X-ray binaries have been observed since the 80s (van der Klis et al. 1987; Miyamoto et al. 1988; Miyamoto et al. 1993). Lags associated to type-C QPOs in BHBs have been measured with RossiXTE (see e.g. Reig et al. 2000; Remillard et al. 2002; Casella et al. 2004; Mu\~noz-Darias et al. 2010).
Van den Eijnden (2017) studied type-C QPOs from a sample of sources observed with RossiXTE and discovered a dependence on source inclination.
In the past few years, new results were obtained with data from other X-ray missions, namely AstroSat, NICER and HXMT. Belloni et al. (2020) used NICER data to obtain the first time lags for type-B QPOs down to energies below 2 keV, an energy range unreachable by the RossiXTE instruments, and showed that in that band variations lag those at higher energies, with a lag increasing with decreasing energy. This result was interpreted by Garc\'\i a et al. (2021) in light of a time-dependent Comptonization model with feedback developed by Karpouzas et al. (2020). The model has been applied to type-C QPOs (see below) and has been very successful in fitting the spectral-timing characteristics and evolution, in particular for data from the bright transient GRS 1915+105 (see Fender \& Belloni 2004 for a review on the source). The model is based on the effects of the feedback between a Comptonizing corona and the accretion disk, which is shown to lead to an oscillation whose spectral and timing features are successfully fitted to the data (see Bellavita et al. 2022). The main limitation of the model (at least for now) is that it assumes that the corona is spherical with constant optical depth and electron temperature. The model does not address the origin of the oscillation (see Mastichiadis et al. 2022), and is able to explain the observed features giving information on the geometry of the accreting flow. The results of the application of this model can be found in Karpouzas et al. (2021), M\'endez et al. (2022), Zhang et al. (2022a,b), Garc\'\i a et al. (2022), Rawat et al. (2023) and Peirano et al. (2023).

The results obtained on GRS 1915+105 with RossiXTE were based on a few hundred observations spanning the full 16-year lifetime of the mission and showed a clear evolution of the properties of the type-C QPO and its lag spectrum (Zhang et al., 2020; M\'endez et al. 2022; Garc\'\i a et al. 2022). While the observed correlations are solid and the interpretation with the model points toward an evolution of the corona and the jet in the system, the evolution on time scales shorter than days has not been studied yet.
In this paper, we present the results of the analysis of an observation of GRS 1915+105 made with AstroSat  that contains an interval of state-C (see Belloni et al. 2000 for a description of the states of GRS 1915+105), with a type-C QPO that changed frequency between $\sim$6.6 and $\sim$2.5 Hz in a relatively short period of several hours. We analyze the properties of the QPO and find a similar evolution, to which we applied the same model.

\section{Observations and data analysis}
\label{sec:observations}

AstroSat (Agrawal 2006; Singh et al. 2014) is an astronomical satellite launched by the Indian Space Research Organization (ISRO) on 28 September 2015 into a 97.6 min low-Earth orbit with $6^{\circ}$ inclination. It includes five scientific instruments: a large-area X-ray instrument (LAXPC, $3-80$ keV), a soft X-ray telescope (SXT, $0.3-8$ keV), a coded-mask hard X-ray instrument (CZTI, $25-150$ keV), an all-sly monitor (SSM, $2.5-10$ keV) and an ultraviolet telescope (UVIT, 130--180 nm, 200--300 nm, 320-5-50 nm). 
The Large-Area X-ray Proportional Counter (LAXPC) is an X-ray proportional counter array covering the energy range $3-80$ keV. It consists of three detectors (referred to as LX10, LX20 and LX30 respectively), with a combined effective area of 6000 cm$^2$ (Yadav et al. 2016a; Antia et al. 2017). At the time of the observation, all three detectors were active, although with different responses. The timing resolution of the LAXPC is 10 $\mu$s with a dead time of 42 $\mu$s. For the full observation, all information about single photons is available. The characteristics of the LAXPC make it very effective for fast-timing measurements.

\begin{table*}
	\centering
	\caption{Log of the eight data intervals analysed here. The columns are: start and end times (seconds from MJD 57451.47437), number of averaged PDS, QPO frequency and FWHM (both in Hz), and statistical significance (in $\sigma$). QPO phase lags in radians are given for the 15--20 keV band with respect to the 3--4 keV band.
	}
	\label{tab:intervals}
	\begin{tabular}{lccccccc} % four columns, alignment for each
		\hline
		Int. & T$_{s}$& T$_{e}$&N$_{PDS}$ & $\nu_{QPO}$& $\Delta_{QPO}$ & $n_\sigma$ & phase lags (rad)\\
		\hline
		A &     0  &   918  & 14 & $5.67\pm 0.03$ & $0.70\pm 0.10$ & 10.2 & $-0.69\pm 0.04$\\
		B &  3235  &  4087  & 13 & $6.61\pm 0.03$ & $0.67\pm 0.09$ & 10.7 & $-0.98\pm 0.05$\\
		C &  6258  &  6717  &  7 & $4.60\pm 0.02$ & $0.45\pm 0.06$ & 12.1 & $-0.50\pm 0.05$\\
		D &  9080  & 10391  & 20 & $4.50\pm 0.01$ & $0.50\pm 0.03$ & 27.7 & $-0.48\pm 0.03$\\
		E & 14925  & 16629  & 27 & $3.54\pm 0.01$ & $0.46\pm 0.02$ & 36.9 & $-0.29\pm 0.02$\\
		F & 20770  & 22933  & 33 & $2.55\pm 0.01$ & $0.32\pm 0.01$ & 40.2 & $-0.11\pm 0.02$\\
		G & 26615  & 29171  & 39 & $2.79\pm 0.01$ & $0.36\pm 0.01$ & 80.8 & $-0.20\pm 0.02$\\
	    H & 32460  & 35475  & 46 & $3.22\pm 0.01$ & $0.41\pm 0.01$ & 47.9 & $-0.19\pm 0.02$\\
		\hline
	\end{tabular}
\end{table*}

\begin{figure}
	\includegraphics[width=\columnwidth]{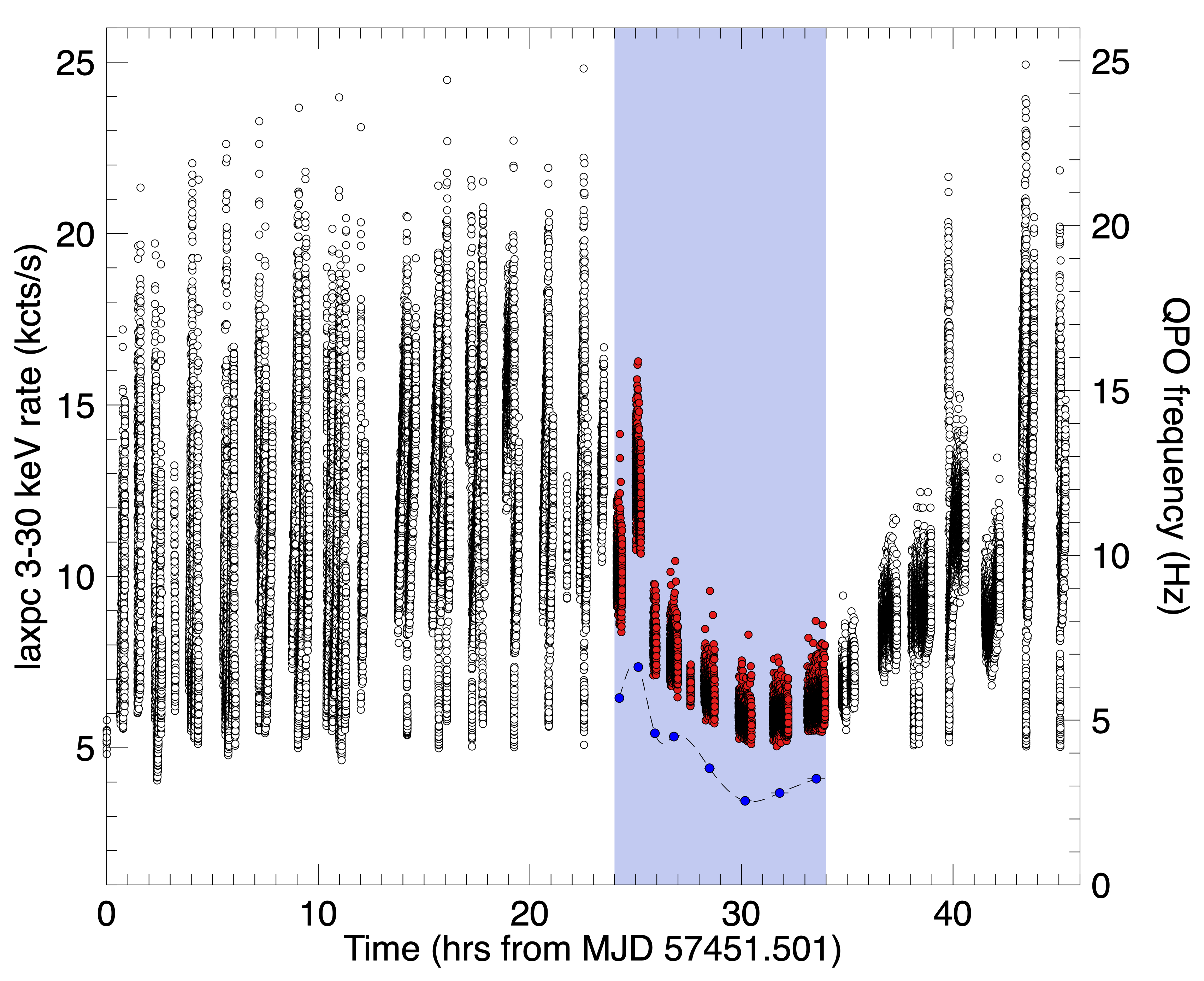}
    \caption{Light curve ($3-50$ keV) for the full LAXPC observation with a time resolution of 1 second. The light blue area and the red data points mark the region considered in this work. 
    The blue points are the QPO centroid frequencies measured for each of the eight intervals considered here (see text).
             }
    \label{fig:licu}
\end{figure}

We analyzed the AstroSat/LAXPC data of GRS 1915+105 taken on 4--6 March 2016 (see Yadav et al. 2016b, where the same observation is analysed) using the GHATS analysis package\footnote{\url{http://astrosat-ssc.iucaa.in/uploads/ghats_home.html}}. The observations span the period from 2016 March 04 11:22 UT to 2016 March 6 08:43 UT for a total net exposure of 63222 s. We show the full $3-50$ keV light curve for this observation in Fig. \ref{fig:licu}. Strong variability typical of GRS 1915+105  (Belloni et al. 2000) is visible, but there is a long period when this structured variability is absent and the count rate follows a `U'-shaped evolution. 
Such an interval has been repeatedly observed from GRS 1915+105 and is one of its `trademarks' (Belloni et al. 1997, 2000).
A preliminary examination of this part of the light curve confirms that this interval is characterized by fast timing properties compatible with state C (associated to variability class $\chi$, Belloni et al. 2000), as previously shown in Yadav et al. (2016). In order to study the properties of the low-frequency QPO observed in state C, we limited our analysis to this part of the observation. The last two intervals (corresponding to AstroSat separate orbits) were affected by data drop outs and were excluded. This resulted in the selection of the region marked in light blue in Fig. \ref{fig:licu}, which corresponds to nine intervals, separated by gaps due to the orbital constraints of the satellite. Since the fifth interval was very short (96 seconds), eight intervals were finally considered, which we named with capital letters from A to H. An observation log can be seen in Tab. \ref{tab:intervals}.

For each of the eight intervals A--H, we produced a Fourier Transform by extracting a light curve with 1 ms time binning and dividing it in segments of duration 65.536 seconds (the number of segments for each interval, $N_{PDS}$, can be seen in Tab. \ref{tab:intervals}). The PDS produced from each segment were then averaged to obtain a single PDS per interval, with a Nyquist frequency of 500 Hz and a lowest frequency of 15.3 mHz.
This procedure was applied to data for the full energy range ($3-50$ keV) as well as for 14 non-overlapping energy bands, shown in Tab. \ref{tab:bands}. From each Fourier Transform, we produced a Power Density Spectrum (PDS), normalized according to Leahy et al. (1983). We fitted the full-band PDS for each interval with a combination of Lorentzian components (Belloni et al. 2002) plus a constant component to account for Poissonian noise. 

\begin{figure}
	\includegraphics[width=\columnwidth]{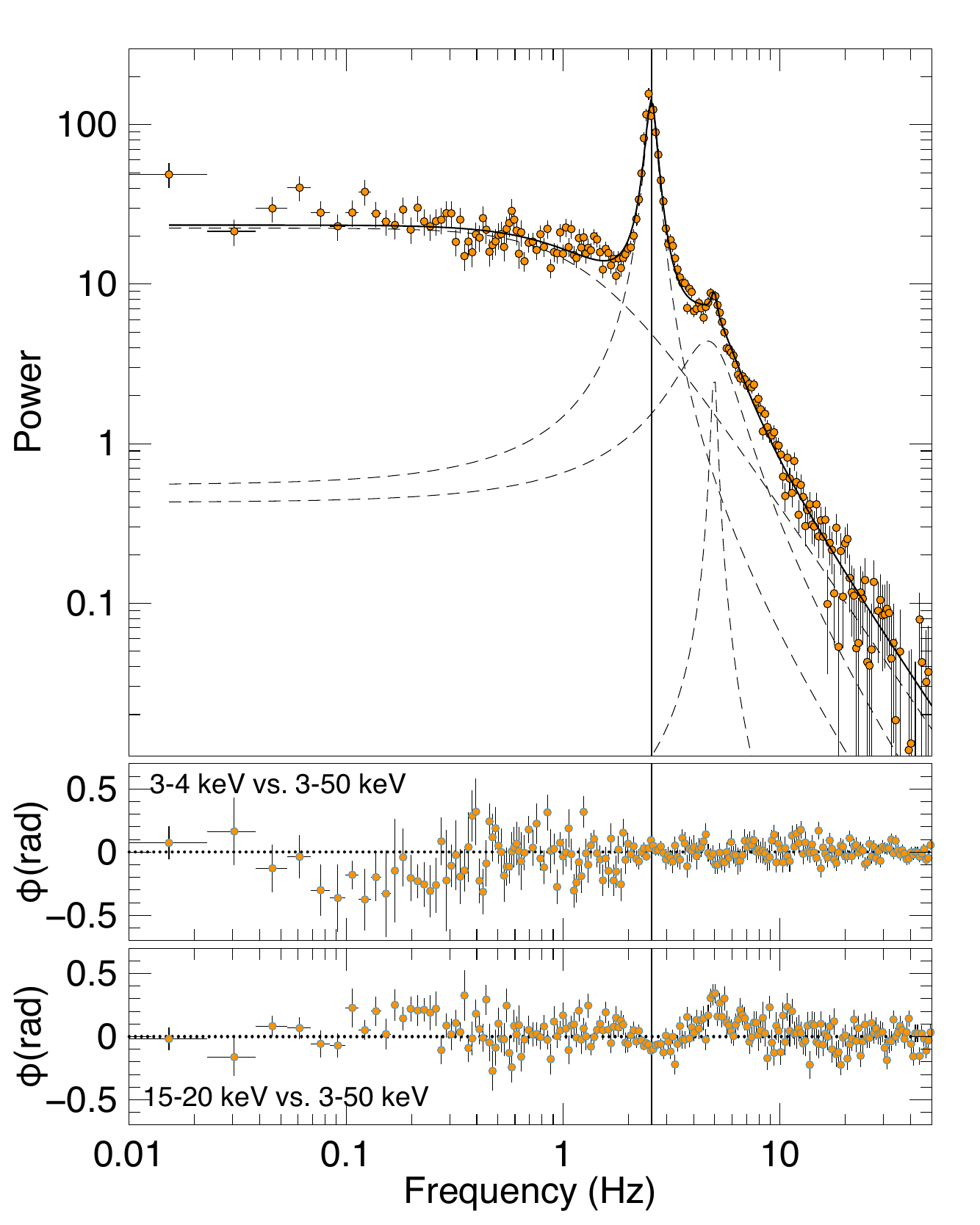}
    \caption{Top panel: PDS of interval F, corresponding to the QPO with the lowest centroid frequency (2.55 Hz, marked by a vertical line). The dashed lines indicate the individual Lorentzian components used to fit the PDS, and the full line their combination.
    Middle panel: phase-lag spectrum between the $3-4$ keV energy band and the full energy band.
    Bottom panel: phase-lag spectrum between the $15-20$ keV energy band and the full energy band.
             }
    \label{fig:pdsf}
\end{figure}

\begin{figure}
	\includegraphics[width=\columnwidth]{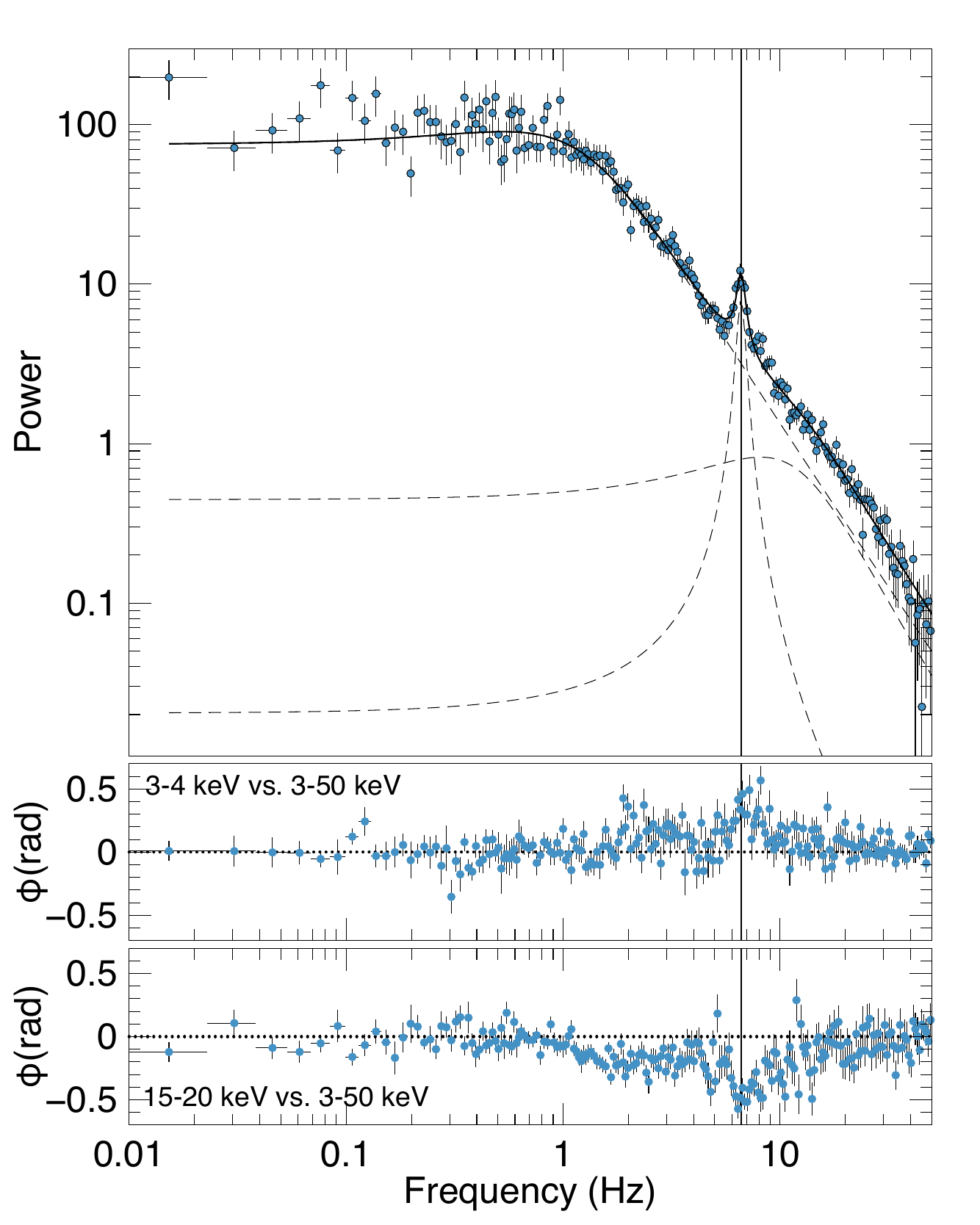}
    \caption{Same as Fig. \ref{fig:pdsf}, but for interval B, corresponding to the QPO with the highest centroid frequency (6.61 Hz).
             }
    \label{fig:pdsb}
\end{figure}

For each of the eight intervals we also used the Fourier Transforms to compute average cross-spectra, using the full $3-50$ keV band as reference (Uttley et al. 2014; Ingram 2019). From the cross-spectra, we
computed the phase lags at the QPO frequency by averaging the real and imaginary parts over a frequency band centred on the centroid frequency and one FWHM wide. As the reference band $3-50$ keV included the photons of all other bands, there is a zero-lag correlation for all cross-spectra, which we removed by estimating its amplitude from the high-frequency part of the cross-spectra, where no source signal is observed. 

\begin{figure}
	\includegraphics[width=\columnwidth]{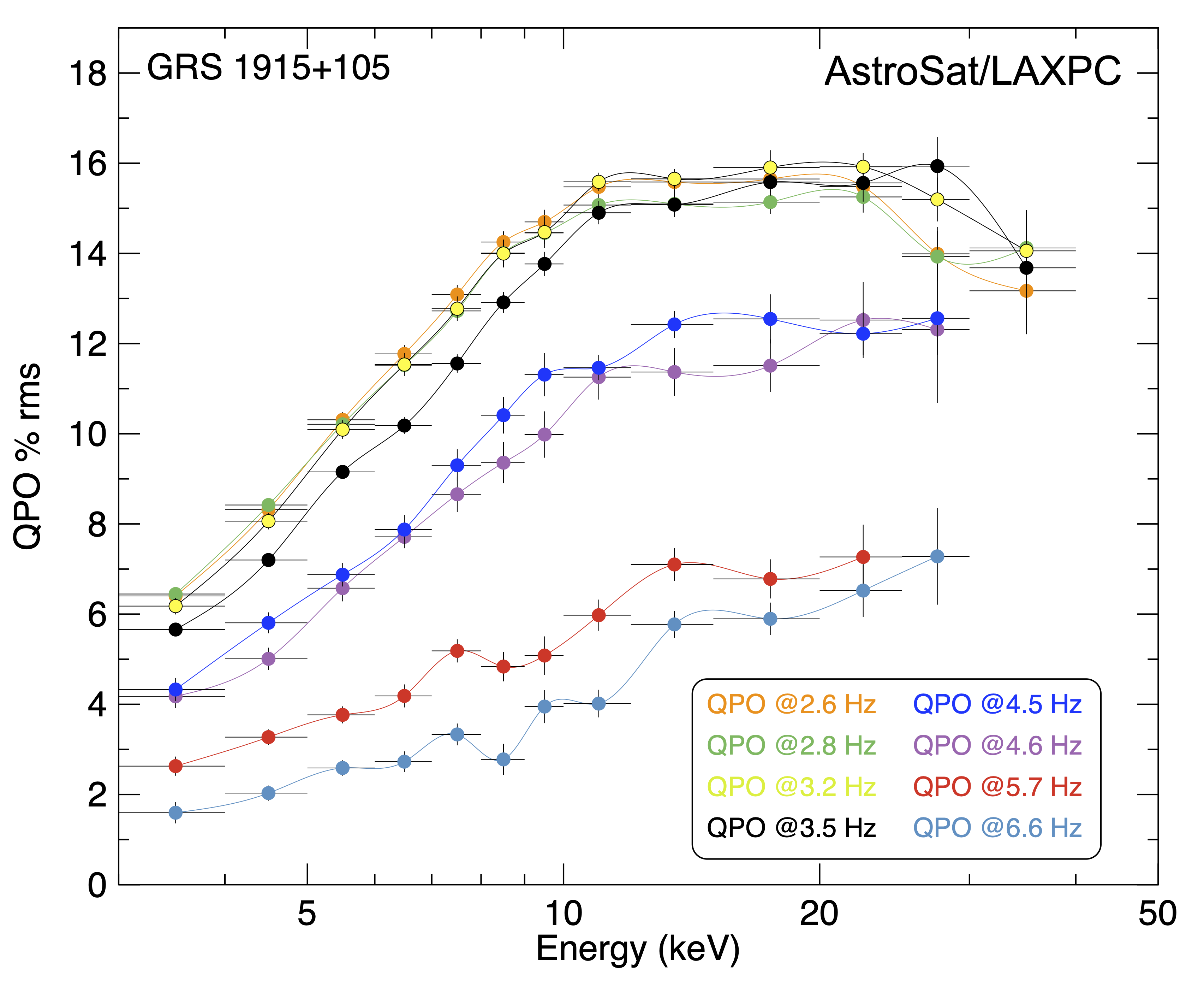}
    \caption{Fractional rms of the QPO as a function of energy for the eight intervals described in the text.
             }
    \label{fig:rms}
\end{figure}

Two examples of PDS, corresponding to the intervals with the QPO at the lowest and highest frequency are shown in the top panels of Figs. \ref{fig:pdsf} and \ref{fig:pdsb}. The centroid frequency of the QPO for all intervals are reported in Tab. \ref{tab:intervals} and plotted in Fig. \ref{fig:licu}. From the best-fitting model, for each PDS we computed the integrated fractional rms amplitude (Belloni \& Hasinger (1990) of the QPO component, which we show in Fig. \ref{fig:rms}. The QPO was not significant in the $40-50$ keV band for all intervals, while for other intervals significant detections went up to 30 keV. In Fig. \ref{fig:rms} we only include the rms amplitude of the QPO in the energy bands in which it was significant, as the upper limits in the remaining bands are not stringent.

\begin{figure}
	\includegraphics[width=\columnwidth]{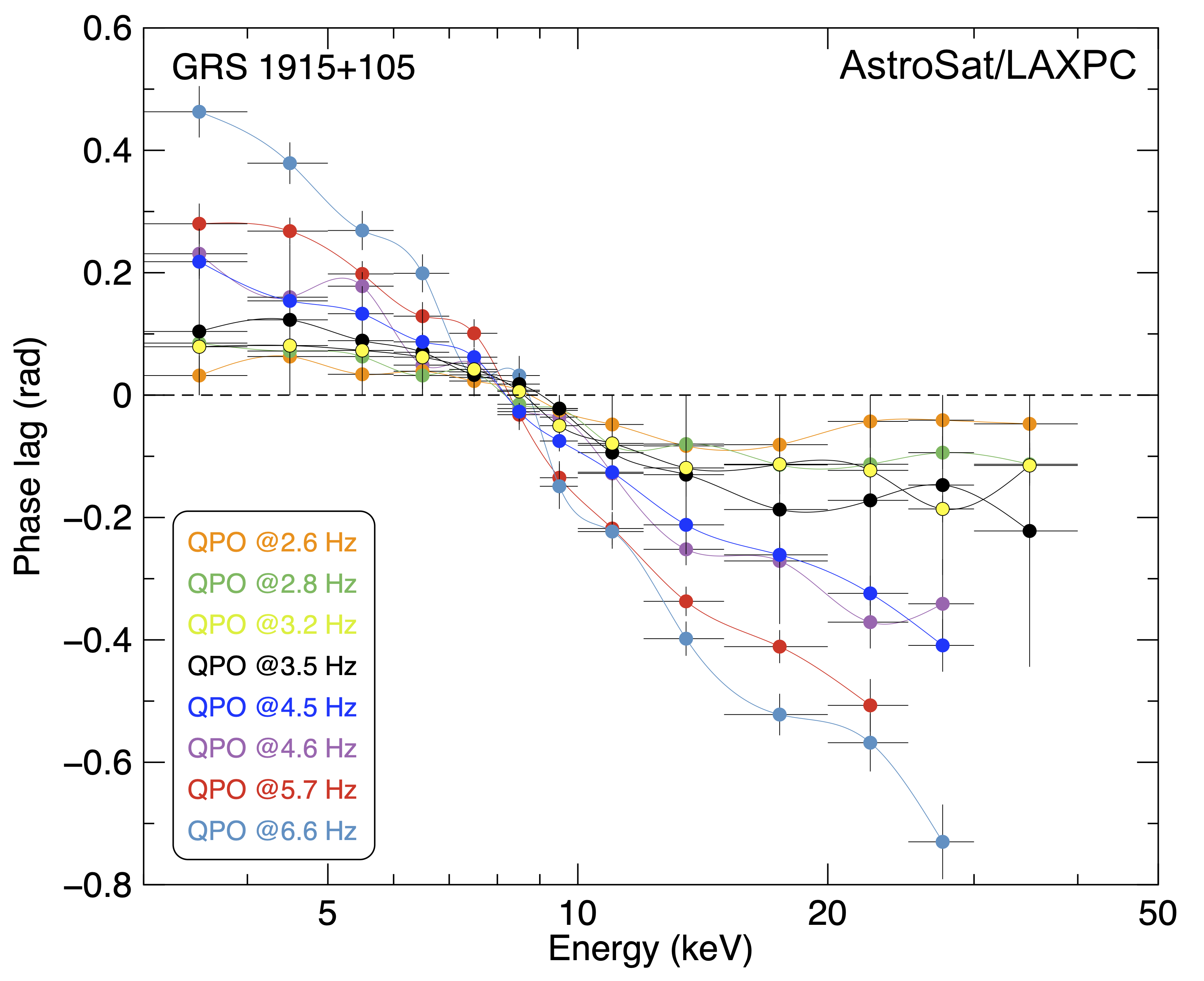}
    \caption{Phase lags of the QPO (in radians) as a function of energy for the eight intervals described in the text. The reference band is the total one (T), $3-50$ keV. 
             }
    \label{fig:lags}
\end{figure}

The eight phase-lag spectra for the separate intervals are shown in Fig. \ref{fig:lags}. As in the case of the fractional rms, we do not include upper limits for the bands in which the QPO was not detected significantly.

The main results of the analysis are shown in Fig. \ref{fig:rms} and \ref{fig:lags}. The fractional rms of the QPO grows with energy until $\sim$10 keV, then flattens and possibly decreases above 30 keV.
The QPO phase lags all decrease with energy, but with a very different slope, which increases as the QPO frequency increases. The lags all have the full band ($3-50$ keV) as reference band, which sets the zero level. The fact that all lag spectra cross zero around 8 keV is most likely due to a combination of the QPO rms spectra and LAXPC response. From Fig. \ref{fig:pdsf} one can see that there is a positive-lag peak corresponding to the second harmonic of the QPO, but the analysis of harmonics goes beyond the scope of this paper.

We extracted LAXPC energy spectra from all the eight intervals in order to check the evolution of Comptonization parameters with QPO frequency.

To obtain the energy spectra, we started from level1 files from the AstroSat archive and obtained level2 products by using format (A) software\footnote{\url{http://astrosat-ssc.iucaa.in/laxpcData}}. We then used the same software to extract a spectrum for each of the two intervals, using only unit20 as it is the best calibrated of the three units (all active at the time of observation). We extracted background spectra using the same time intervals. A 2\% systematic error was added to the spectra.

Instead of fitting only the energy spectra, we applied the model by
Bellavita et al.(2022) to the energy and timing data. Specifically, for
each interval, following Garc\'\i a et al. (2022), we fitted simultaneously the average energy spectrum of the source (in the $3-40$ keV energy range), and the rms the phase lag spectrum of the QPO. 

In XSPEC parlance, the model used to fit the average energy spectrum is {\tt PHABS (DISKBB + NTHCOMP)}, where {\tt PHABS} is the interstellar absorption (fitted parameter is $N_H$), {\tt DISKBB} a thermal disk component (fitted parameters are the inner disk temperature $kT_{in}$ and the normalization $N_{dbb}$ proportional to the square of the inner radius), and {\tt NTHCOMP} a thermal Comptonization model (parameters are the photon index $\Gamma$, the electron temperature $kT_e$ and a normalization, Zdziarski, Johnson \& Magdziarz 1996; Zycki, Done \& Smith 1996). The temperature of the seed photons in the  {\tt NTHCOMP} model was linked to the accretion-disk inner temperature, $kT_{in}$. The value for interstellar absorption was left free, but tied between the eight intervals, under the assumption that any local absorption would not vary on such a short time scale.

For the lag and rms spectra, we fitted the model {\tt VKOMPTHDK * DILUTION}, where {\tt VKOMPTHDK}\footnote{\url{https://github.com/candebellavita/vkompth}} is the variable-Comptonization spectral-timing model developed by Bellavita et al. (2022). Its parameters are the same $\Gamma$ and $kT_e$ of {\tt NTHCOMP}, plus the cloud size $L$, the feedback fraction $\eta$, the external heating $\delta \dot H_{\rm ext}$ and the reference lag {\it reflag}. {\tt DILUTION} is a multiplicative model to take into account the fact that the fractional rms of the (variable) Comptonization component would be diluted by the (constant) contribution from another non-variable component, in this case, the thermal accretion disk. The dilution factor was fixed to 1 for the lag spectrum (where there is no dilution) and to the ratio of contribution of the corona component and the total one for the rms spectrum. The spectral parameters $kT_{in}$, $\Gamma$ and $kT_e$ of the {\tt DISKBB} and {\tt NTHCOMP} spectral components were tied to the same parameters of the {\tt VKOMPTHDK} component used for rms and lag spectra. In this way, it was possible to fit simultaneously the average energy spectrum of the source and the spectra of the timing parameters of the QPO.
Notice that Garc\'\i a et al. (2022) used a blackbody model for the accretion disk emission, while we adopted an improved version of the model that assumes a disk-blackbody as the disk-emitting component.

\begin{table}
	\centering
	\caption{Energy boundaries (in keV) for the full range ($T$) and the fourteen energy bands (1 to 14) used in the analysis.
	}
	\label{tab:bands}
	\begin{tabular}{lcc} 
		\hline
		Band & E$_{low}$ & E$_{high}$\\
		\hline
		T  &  3 & 50\\
		1  &  3 &  4\\
		2  &  4 &  5\\
		3  &  5 &  6\\
		4  &  6 &  7\\
		5  &  7 &  8\\
		6  &  8 &  9\\
		7  &  9 & 10\\
	    8  & 10 & 12\\
	    9  & 12 & 15\\
	    10 & 15 & 20\\
	    11 & 20 & 25\\
	    12 & 25 & 30\\
	    13 & 30 & 40\\
	    14 & 40 & 50\\

		\hline
	\end{tabular}
\end{table}

%\section{Results}
\label{sec:results}

\begin{figure}
	\includegraphics[width=\columnwidth]{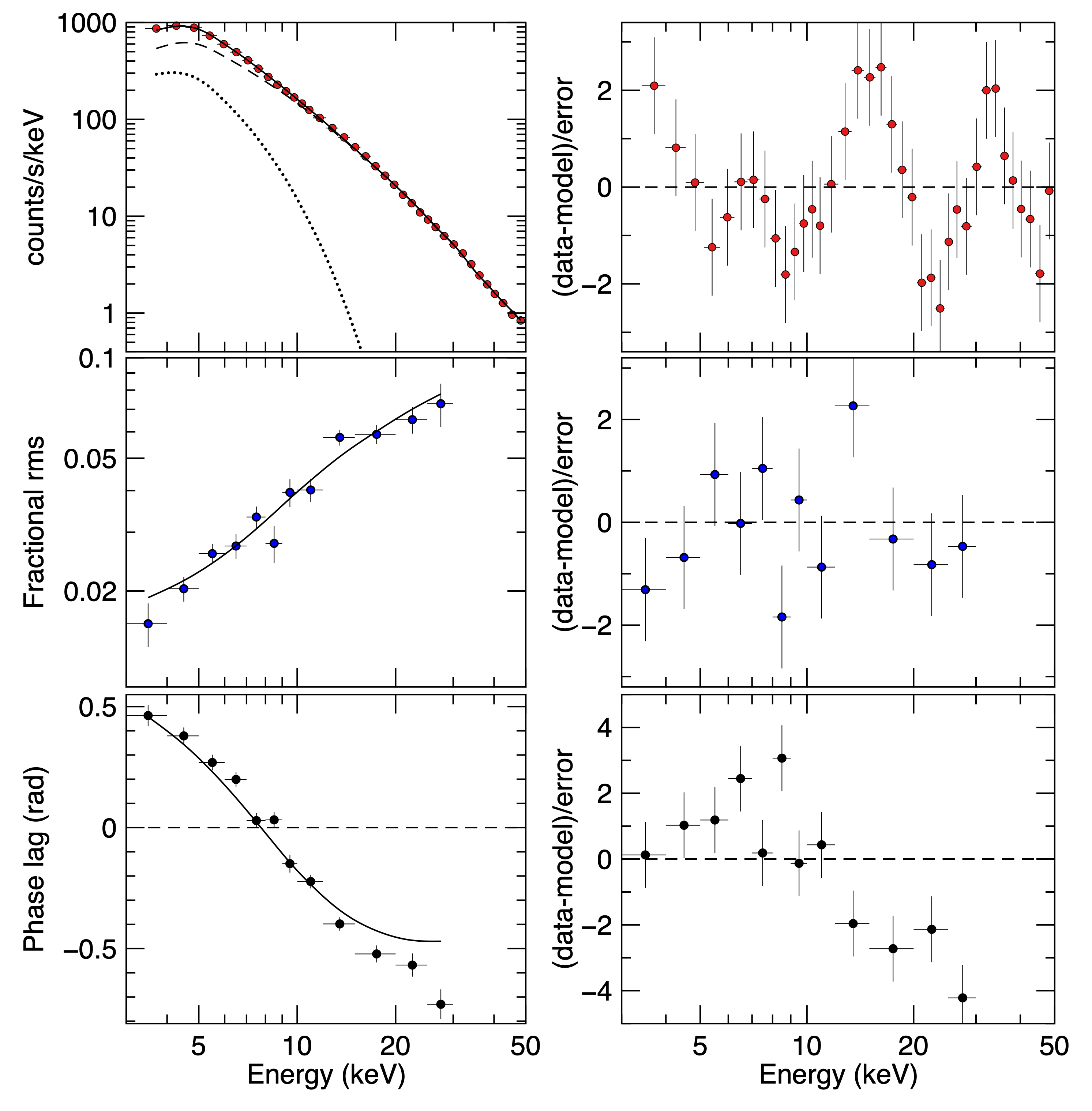}
    \caption{Left column: the best fits to flux, rms and phase-lag spectra for interval B. In the top panel the curves for the two spectral components (disk and Comptonization) are also shown. Right column: corresponding residuals.
             }
    \label{fig:fit}
\end{figure}

The best-fitting parameters are shown in Tab. \ref{tab:spectra}. The joint best fit value for the interstellar absorption was $5.3(6)\times 10^{22}$ cm$^{-2}$ The model represents the data reasonably well as far as the main trend is concerned, although the chi-square values indicated that statistically the fits are not good, in particular as the highest-energy points in the rms and lag spectra show deviations. An example of best fit for the case of interval B, with residuals, is shown in Fig. \ref{fig:fit}.
A full detailed analysis of the spectra is beyond the scope of this work and would require moving to the dual model by Garc\'\i a et al. (2021): the current fits are sufficient to identify the changes at different QPO frequencies.
It is clear that the as the QPO frequency increases from 2.55 Hz (interval F) to 6.61 Hz (interval B), the parameter that changes the most is the size of the corona, which increases by a factor $\sim$5 (see Fig. \ref{fig:parameters}, middle panel). At the same time the power-law index $\Gamma$ increases as usually observed for type-C QPOs (Motta et al. 2009, see bottom panel in Fig. \ref{fig:parameters}), and the accretion disk becomes hotter and its inner radius increases.

The best-fitting values for the feedback fraction $\eta$ are similar across the intervals. The parameter $\eta$ represents the fraction of disk flux due to the photons from the corona that return to the disc. Such high values imply that for both intervals the disk flux is almost entirely due to feedback. From the best-fitting values, we can also estimate the fraction $\eta_{int}$ of photons emitted by the corona that return to the disk (see Karpouzas et al. 2020 for a more detailed description), which is shown in the top panel of Fig. \ref{fig:parameters}. Ignoring relativistic effects, an increase in the solid angle subtended by the disk as seen by the corona will lead to an increase in $\eta_{int}$, which is qualitatively what is observed. A more quantitative comparison would require a much more detailed model.

\section{Discussion}
\label{sec:discussion}

We have found an AstroSat observation of GRS 1915+105 in which the source went through one of the typical `U'-shaped events, corresponding to state C, when the type-C QPO is detected in this source. However, while in most of the cases these events are either short, below $\sim$1 hr (Belloni et al. 1997; Markwardt, Swank \& Taam 1999; Belloni et al. 2000), or very long, days to months (Fender et al. 1999; Trudolyubov 2001; M\'endez et al. 2022), here the full `U' takes about ten hours. Therefore, it was possible to follow at high signal-to-noise (also due to the high flux of the source and large effective area of the LAXPC) the evolution of the rms and lag spectra of the type-C QPO. As a function of increasing energy, the QPO rms increases, then flattens around $\sim$10 keV, with a possible decrease above $\sim$30 keV (see Fig. \ref{fig:rms}). Again as a function of increasing energy, the QPO lag spectrum decreases with a slope that is higher for QPOs with a higher centroid frequency (see Fig. \ref{fig:lags}). Remarkably, as a function of QPO frequency the parameters of the corona change much more significantly than those of the disc (Table~\ref{tab:spectra}). Together with the significant changes of the QPO rms and lag spectra with QPO frequency, this bolsters the idea that the properties of the QPO are driven, predominantly, by the corona, as proposed in the time-dependent Comptonisation model (Karpouzas et al. 2020; Bellavita et al. 2022).

\begin{figure}
	\includegraphics[width=\columnwidth]{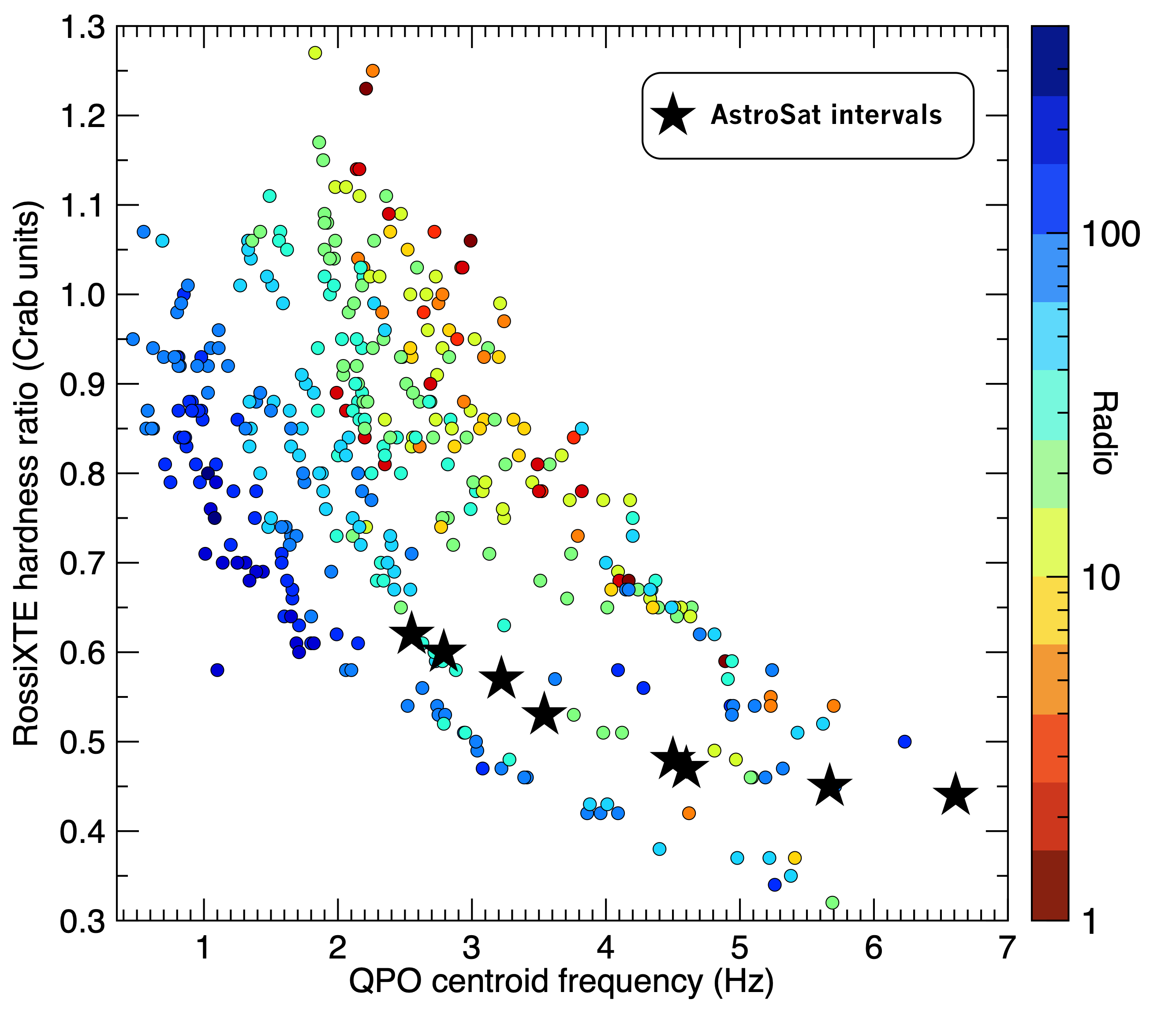}
    \caption{Plot of hardness ratio versus QPO frequency for RossiXTE observations of GRS 1915+105 (from M\'endez et al. 2022), with radio flux (in mJy) coded in the symbol colors. The black stars represent our eight AstroSat intervals, for which we have no radio information. The RossiXTE hardness ratios have been computed using our best-fit spectral models and the response of the PCA instrument.
             }
    \label{fig:radio}
\end{figure}

\begin{figure}
	\includegraphics[width=\columnwidth]{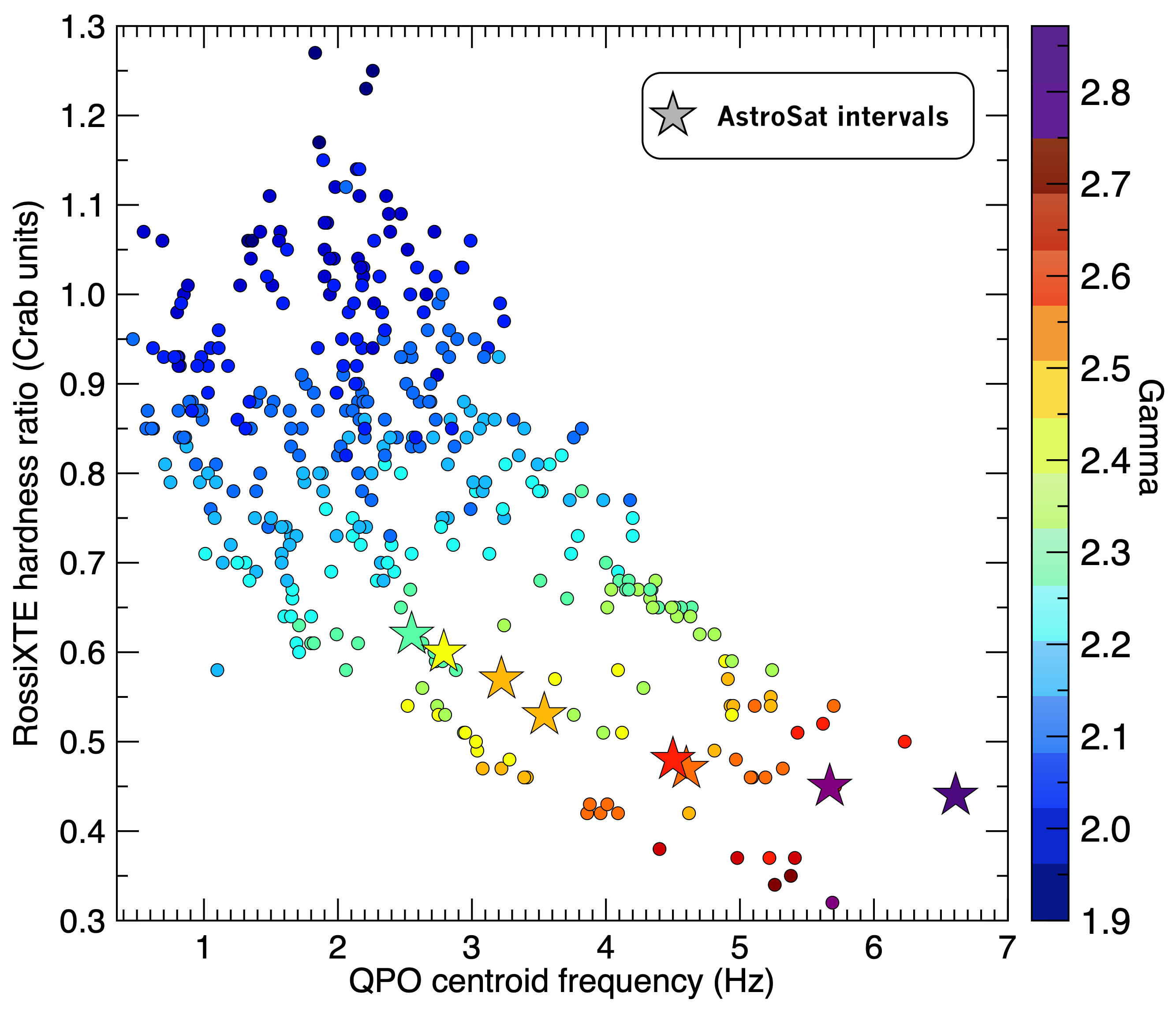}
    \caption{Same as Fig . \ref{fig:radio}, with spectral index $\Gamma$ coded in the symbol colors. The stars represent our eight AstroSat intervals, with the same color coding according to our best-fit $\Gamma$ values.
             }
    \label{fig:gamma}
\end{figure}

The observation reported here was analyzed by Yadav et al. (2016b). In that work, nine time intervals are included, although no light curve is shown and the times do not seem to match ours. Three time-lag and rms spectra are shown in their Fig. 9. From the QPO centroid frequency they appear to correspond to our intervals F, D and B, although this identification does not match the light curves. Strangely, their lag spectra extend to 70 keV, while the rms spectra only reach 35 keV. Comparing the lag spectra we can see that they are compatible with ours, although the one for the 2.55 Hz QPO has a smaller amplitude. The rms spectra show a drop above 25 keV not present in ours and overall are lower by about 60\%. The description of their analysis is not sufficiently detailed to identify the reasons for the differences. Only three cases are presented and no interpretation or model application is made in that work.

The results we found both for the phase lags and the rms are consistent with those reported by Zhang et al. (2020). However, Zhang et al. (2020) analyzed 620 RossiXTE/PCA observations when GRS 1915+105 was in its state C, irregularly sampled and covering the full operational life of RossiXTE, 16 years. Based on those results, Karpouzas et al. (2021) and Garc\'\i a et al. (2022) applied the Comptonization model developed by Karpouzas et al. (2020) to the RossiXTE/PCA data, and together with the radio measurements, M\'endez et al. (2022) proposed a scenario for the accretion/ejection geometry in this system: At high QPO frequency, $\sim 6-8$ Hz, the QPO lags are soft and the magnitude of the lags is the largest, the X-ray corona is large (size $\approx 2000$ km) and relatively hot ($kT_e \gtrsim 15-20$ keV), and the radio emission from the jet is low ($\lesssim 5$ mJy; see Fig. 4 and Extended Data Fig. 4 in M\'endez et al. 2022). As the QPO frequency decreases, the magnitude of the lags decreases (the lags continue to be soft), $kT_e$ and the size of the X-ray corona decrease and the radio flux increases. When the QPO frequency crosses below $\sim 2$ Hz the lags become hard and their magnitude increases with decreasing QPO frequency, the corona is the coolest ($kT_e \lesssim 5-6$ keV) and the jet emission is the strongest ($\gtrsim 100$ mJy). Based on that, M\'endez et al. (2022) proposed that the X-ray corona morphs into the radio jet and vice versa.

\begin{figure}
	\includegraphics[width=\columnwidth]{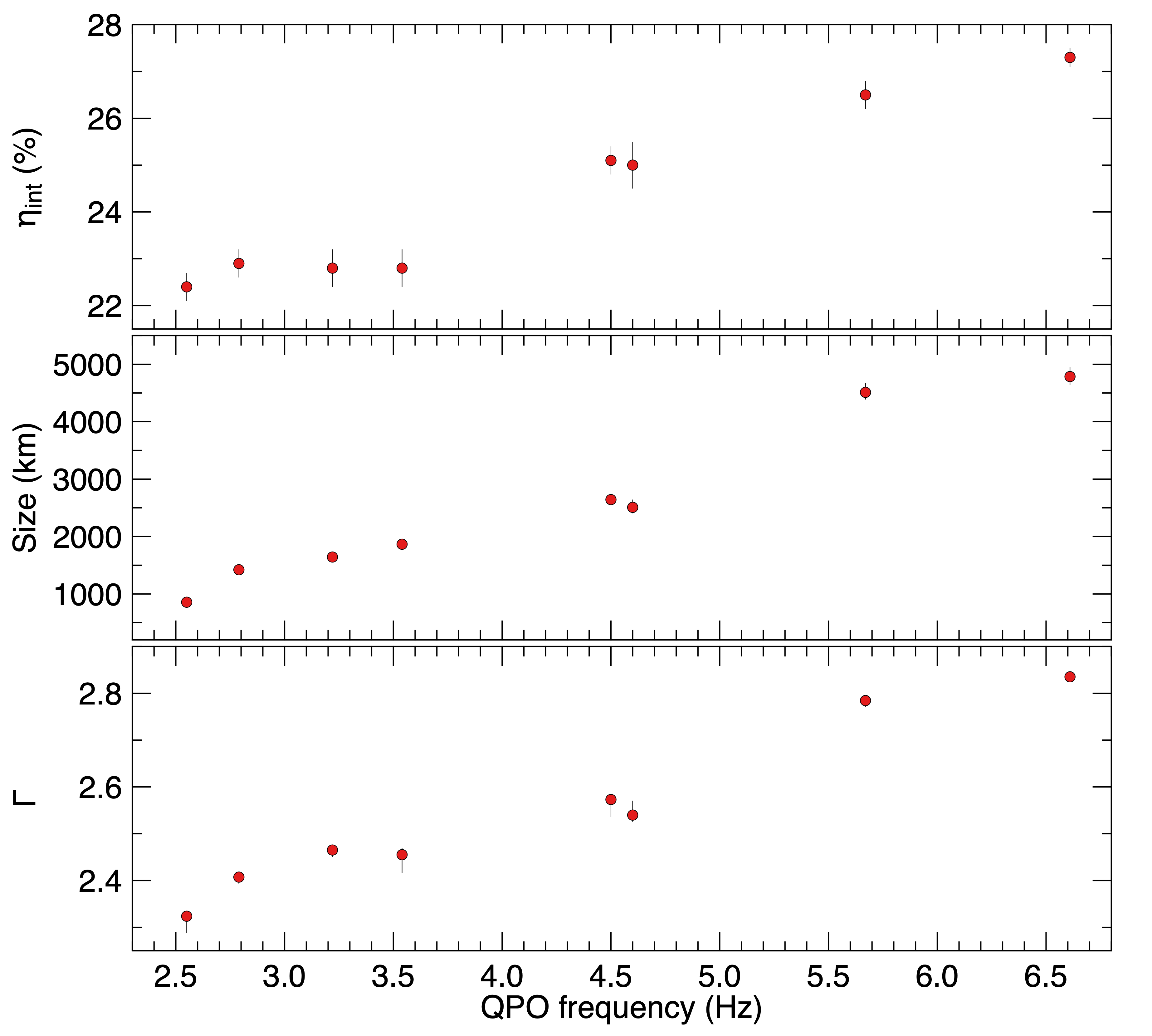}
    \caption{Plot of selected model parameters versus QPO frequency for the eight intervals. From top to bottom: $\eta_{int}$, representing the fraction of photons emitted by the corona that return to the disk (this is not a direct fit parameter, but it is derived from the others), the size of the corona in km and the spectral index $\Gamma$.
             }
    \label{fig:parameters}
\end{figure}

The evolution of the spectral/timing parameters in these AstroSat observations is consistent with what was observed by M\'endez et al. (2022), whose work was based on the analysis of a few hundred observations of GRS 1915+105 with type-C QPOs made with RossiXTE. Their observations spanned the full operational life of the satellite, more than 15 years, and therefore sampled more or less randomly many events in the time evolution of the source. Here the evolution of the disk and the corona is observed in a period as short as 11 hours, showing that the same behaviour takes place during a single event.
Even the evolution of the size of the corona with QPO frequency is the same (see their Extended Data Fig. 4), increasing by a factor of $\sim$5 over the observed range in QPO frequencies (there is an offset in absolute values, which is likely caused by the difference in energy range between the RossiXTE and AstroSat instruments). 
In this case the QPO frequency does not reach down to 2 Hz and the lags do not switch from soft to hard. Since there was no simultaneous radio coverage during the AstroSat observations, we cannot compare the properties of the jet in our data and those of RossiXTE. 
 
Because the RossiXTE observations were random pointings over the long time evolution of the source whereas our observations cover an event lasting several hours, our results confirm that the size of the corona is linked to the QPO frequency and not dependent on the time scale of the events.
Intervals of this type have been observed on shorter time scales, from one hour (Belloni et al. 1997; Markwardt et al. 1999), down to $\sim$10 s (variability class $\mu$, Belloni et al. 2000). It is natural to speculate that the same qualitative variations of the corona and the disk take place also for these short-duration events, but the analysis of those requires a much higher signal and is something that will be explored with future missions such as eXTP.

\begin{table*}
	\centering
	\caption{Best-fitting parameters for all eight intervals.
	}
	\label{tab:spectra}
	\begin{tabular}{lcccc} 
		\hline
		Parameter                         & Interval A        &Interval B         & Interval C        & Interval D         \\
		\hline
        QPO $\nu_0$ (Hz)                  & $5.67$            & $6.61$            & $4.60$            & $4.50$             \\
        \hline
	$N_{\rm H}~(10^{22}~\mathrm{cm}^{-2})$&\multicolumn{4}{c}{5.3  $\pm 0.6$}    \\
    $kT_{\rm in}~\mathrm{(keV)}          $& 1.199 $_{-0.027}^{+0.033}$ & 1.364 $_{-0.027}^{+0.030}$ & 1.202 $_{-0.021}^{+0.007}$ & 1.197 $_{-0.021}^{+0.007}$  \\
    $N_{dbb}                             $& 316.9 $_{-15.4}^{+25.8}$   & 273.6 $_{-10.2}^{+21.2}$   & 323.8 $_{-9.2}^{+67.4}$    & 300.9 $_{-8.2}^{+30.5}$     \\
    $\Gamma                              $& 2.784 $_{-0.013}^{+0.008}$ & 2.835 $_{-0.012}^{+0.010}$ & 2.540 $_{-0.014}^{+0.031}$ & 2.573 $_{-0.037}^{+0.008}$  \\
    $kT_{\rm e}~\mathrm {(keV)}          $& $> 372$                    & $> 654$                    & 20.4  $_{-1.5}^{+2.8} $    & 28.1  $_{-4.2}^{+0.9} $    \\
    $N_{\rm nthComp}                     $& 3.74  $_{-0.25}^{+0.15}$   & 3.87  $_{-0.21}^{+0.20}$   & 2.31  $_{-0.08}^{+0.11}$   & 2.30  $_{-0.06}^{+0.01}$ \\
    $\mathrm{Size~(km)}                  $& 4510.9$_{-124.3}^{+163.4}$ & 4785.7$_{-145.9}^{+167.8}$ & 2508.1$_{-103.2}^{+135.3}$ & 2642.9$_{-43.2}^{+36.1}$ \\
    $\eta                                $& $> 0.996$                 & $> 0.997$                   & $> 0.987$                  & $> 0.991$          \\
    $\delta \dot H_{\rm ext}             $& 0.079 $\pm$ 0.002         & 0.072 $\pm$ 0.003           & 0.122 $\pm$ 0.003          & 0.125 $\pm$ 0.002  \\
    $\eta_{int} (\%)                     $& 26.5 $\pm$ 0.3            & 27.3 $\pm$ 0.2              & 25.0 $\pm$ 0.5             & 25.1 $\pm$ 0.3     \\
    $\chi^2 (d.o.f.)                     $& 106.7 (50)                & 129.7 (52)                  & 72.8 (52)                  & 106.0 (52)     \\
		\hline\hline
		Parameter                         & Interval E        &Interval F         & Interval G        & Interval H         \\
		\hline
        QPO $\nu_0$ (Hz)                  & $3.54$            & $2.55$            & $2.79$            & $3.22$             \\
        \hline
	$N_{\rm H}~(10^{22}~\mathrm{cm}^{-2})$&\multicolumn{4}{c}{5.3  $\pm 0.6$}    \\
    $kT_{\rm in}~\mathrm{(keV)}          $& 1.232  $_{-0.014}^{+0.001}$ & 1.234 $_{-0.020}^{+0.013}$ & 1.178 $_{-0.009}^{+0.007}$ & 1.198 $_{-0.014}^{+0.002}$  \\
    $N_{dbb}                             $& 250.7 $_{-17.3}^{+43.6}$   & 180.4 $_{-10.4}^{+4.7}$   & 202.3 $_{-32.1}^{+62.0}$    & 241.0 $_{-52.1}^{+48.3}$     \\
    $\Gamma                              $& 2.455 $_{-0.039}^{+0.014}$ & 2.324 $_{-0.036}^{+0.007}$ & 2.407 $_{-0.015}^{+0.005}$ & 2.465 $_{-0.015}^{+0.005}$  \\
    $kT_{\rm e}~\mathrm {(keV)}          $& 18.7  $_{-0.7}^{+0.2} $    & 13.8  $_{-1.3}^{+0.9} $    & 16.7  $_{-0.7}^{+0.4} $    & 20.0  $_{-1.1}^{+0.5} $    \\
    $N_{\rm nthComp}                     $& 1.70  $_{-0.02}^{+0.01}$   & 1.42  $_{-0.14}^{+0.01}$   & 1.67  $_{-0.06}^{+0.01}$   & 1.89  $_{-0.01}^{+0.01}$ \\
    $\mathrm{Size~(km)}                  $& 1865.9$_{-45.6}^{+52.8}$   & 856.2$_{-69.3}^{+88.7}$    & 1421.6$_{-46.7}^{+43.4}$   & 1643.1$_{-36.4}^{+31.8}$ \\
    $\eta                                $& 0.957 $_{-0.023}^{+0.002}$ & $> 0.915$                  & 0.969 $_{-0.009}^{+0.008}$ & 0.940 $_{-0.011}^{+0.008}$ \\
    $\delta \dot H_{\rm ext}             $& 0.146 $\pm$ 0.002         & 0.137 $\pm$ 0.003           & 0.136 $\pm$ 0.001          & 0.142 $\pm$ 0.001  \\
    $\eta_{int} (\%)                     $& 22.8 $\pm$ 0.4            & 22.4 $\pm$ 0.3              & 22.9 $\pm$ 0.3             & 22.8 $\pm$ 0.4     \\
    $\chi^2 (d.o.f.)                     $& 178.5 (54)                & 209.0 (54)                  & 203.4 (54)                 & 267.7 (54)     \\
   \hline
	\end{tabular}
\end{table*}

\section*{Acknowledgements}

Unfortunately, Tomaso Belloni passed away before this paper was accepted. We will miss his insightful ideas about these and other topics. Given that this will likely be his final first-author paper, we would like to let Tomaso express his gratitude to his daughter Alice and his son Alessandro who, as he told us many times, were his most cherished treasures and his utmost source of inspiration.
This work makes use of data from the AstroSat mission of the Indian Space Research Organisation (ISRO), archived at Indian Space Science Data Centre (ISSDC). 
TMB acknowledges financial contribution from grant PRIN INAF 2019 n.15.
MM acknowledges support from the research programme Athena with project number 184.034.002, which is (partly) financed by the Dutch Research Council (NWO). F.G. is a CONICET researcher and acknowledges support from PIP 0113 (CONICET), PICT-2017-2865 (ANPCyT) and PIBAA 1275 (CONICET).
We benefited from discussions during Team Meetings of the International Space Science Institute (Bern), whose support we acknowledge. We would normally thank the referee for their comments on the manuscript. This time, however, this is not appropriate.
%The referee was very, very, very bad!!!!! 

\section{Data availability}
\label{sec:data_availability}

The data used in this article are publicly available at the website of the ISRO Science Data Archive for AstroSat Mission, \url{https://astrobrowse.issdc.gov.in/astro\_archive/archive/Home.jsp}. The {\sc vKompth model is publicly available in a Github repository (\url{https://github.com/candebellavita/vkompth}).}


% Don't change these lines
\bsp	% typesetting comment
\label{lastpage}
\end{document}